\documentstyle[11pt,IAU207_pasp,twoside,psfig]{article}
\def\edcomment#1{\iffalse\marginpar{\raggedright\sl#1\/}\else\relax\fi}
\reversemarginpar

\begin{document}
\title{HST-WFPC2 Observations of the Star Clusters in the Giant HII Regions of M33}
 \author{Myung Gyoon Lee, Hong Soo Park , Sang Chul Kim}
\affil{Astronomy Program, SEES, Seoul National University, Seoul, 151-742, Korea}
\author{William H. Waller}
\affil{Department of Physics and Astronomy, Tufts University, Medford, MA 02155, USA}
\author{Joel Wm. Parker}
\affil{Southwest Research Institute, Suite 426, 1050 Walnut Street, Boulder, Colorado 80302, USA}
\author{Eliot M. Malumuth}
\affil{Raytheon ITSS and Laboratory for Astronomy and Solar Physics, 
NASA/Goddard Space Flight Center, Code 681, Greenbelt, MD 20771, USA}
\author{Paul W. Hodge}
\affil{Astronomy Department, University of Washington, Box 351580, Seattle, WA 98195-1580a, USA}

\begin{abstract}
We present a photometric study of the stars in ionizing star clusters
embedded in several giant HII regions of M33 
(CC93, IC 142, NGC 595, MA2, NGC 604 and NGC 588). 
Our photometry are based on the HST-WFPC2 images of these clusters. 
Color-magnitude diagrams and color-color diagrams of these
clusters are obtained and are used for estimating the reddenings and 
ages of the clusters. 
The luminosity functions (LFs) and initial mass functions (IMFs)
of the massive stars in these clusters are also derived.
The slopes of the IMFs range from $\Gamma = -0.5$ to $-2.1$.
It is found interestingly 
that the IMFs get steeper with increasing galactocentric distance 
and with decreasing [O/H] abundance. 
\end{abstract}


M33, a spiral galaxy in the Local Group, provides an ideal laboratory
to investigate the properties of ionizing star clusters and how much
metal abundance affects stellar population at the high-mass end.
Interestingly the giant H II regions in M33 are not powered by dense
superclusters -- as is the case in 30 Doradus, the Antennae and other
more distant starbursting systems.  
Instead, the clusters underlying the giant HII regions in M33 sprawl 
over ~100pc. Two bright clusters NGC 595 and NGC 604 in M33 were
studied by Malumuth et al. (1996) and Hunter et al. (1996), respectively.

We have analyzed, using the HSTphot package (Dolphin 2000), 
the HST-WFPC2 images which were obtained with four filters: 
F170W, F336W, F439W and F547M for CC93, IC 142, NGC 595, MA2, and NGC 588,  
and with three filters, F336W, F555W and F814W for NGC 604.
Figure 1 shows color-magnitude diagrams of the measured stars in the
clusters. 
It is seen that all the clusters show a prominent blue plume which consists
of blue supergiants and massive main sequence stars.
The blue plumes extend up to $M_V \approx -8$, 
indicating that these clusters are very young.
We have determined the reddenings for these clusters using the color-color diagrams, obtaining mean values of the reddenings of $E(B-V)=0.08 - 0.36$. 


The ages of the clusters are estimated  using the Geneva isochrones 
(Lejeune \& Schaerer 2001).
We have adopted the metallicity of these clusters converted from [O/H] of HII regions given by Vilchez et al. (1988). 
The blue plumes in the clusters in Figure 1 are fit reasonably well 
by the theoretical isochrones with young ages ranging from 3.2 Myrs to 6.3 Myrs.


We have derived the luminosity functions (LFs) of the stars in the blue plume
from the color-magnitude diagrams.
For the estimation of the background contribution, we have selected regions
located in the WF chip fields. 
We also have estimated the completeness of our photometry of each cluster
using the artificial star experiment with the HSTphot package.
The logarithmic slopes of the LFs are found to range 
from $\alpha = -0.4$ to --0.8.
Then the initial mass functions (IMFs) of the massive stars in these clusters
 are derived from the LFs.
The IMFs of these clusters are
 fit reasonably well by the power laws ($\log \xi \propto \Gamma \log M$ ). 
The slopes of the IMFs for the massive stars with $1.0<\log M<1.67$ ($10< M/M_{\odot} <46$) are found to vary significantly from $\Gamma = -0.5$ 
to $-2.1$. %

Figure 2 displays the relations of 
 the slopes of the LFs and IMFs,
the galactocentric distance 
($R/R_0$ where $R_0 = 28'$ is a scale length of the disk in M33), 
and the [O/H] abundance, 
Figure 2 shows good correlations among them:
$\Gamma \propto (-1.64\pm0.41)R$, 
$\Gamma \propto (1.94\pm0.83)[$O/H$]$.
It is found that the IMFs of the clusters get steeper with increasing
radial distance (and with decreasing [O/H] abundance).
This IMF steepening may represent the first strong evidence for 
a systematic environmental effect on stellar population at the high-mass
end. 
The photoevaporative process provides a viable mechanism for ablating
massive protostellar cores and thus steeping the IMF.
This result leads to a prediction that the most top-heavy (flattest)
IMFs may occur near the centers of star-forming galaxies.

\acknowledgments
This research is supported in part by the MOST/KISTEP International Collaboration Research Program (1-99-009). 

\begin{figure}
\centerline{
\psfig{figure=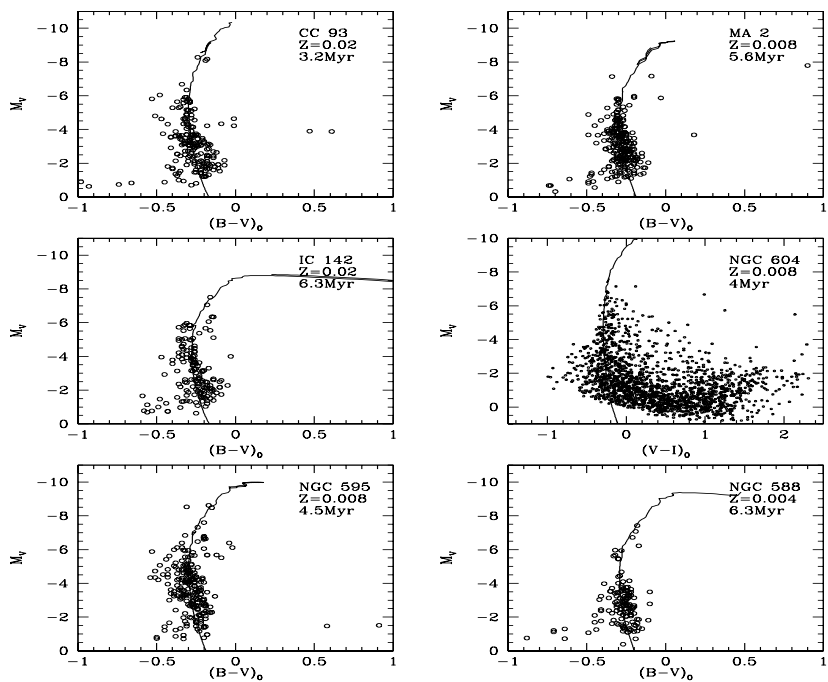,width=12.5cm,height=9cm} }
\caption{Isochrones (solid lines) fit to the data of the clusters in M33.
 Corresponding metallicity and age for each cluster  are given 
in the upper right corner of each panel.}
\end{figure}

\begin{figure}
\centerline{ 
\psfig{figure=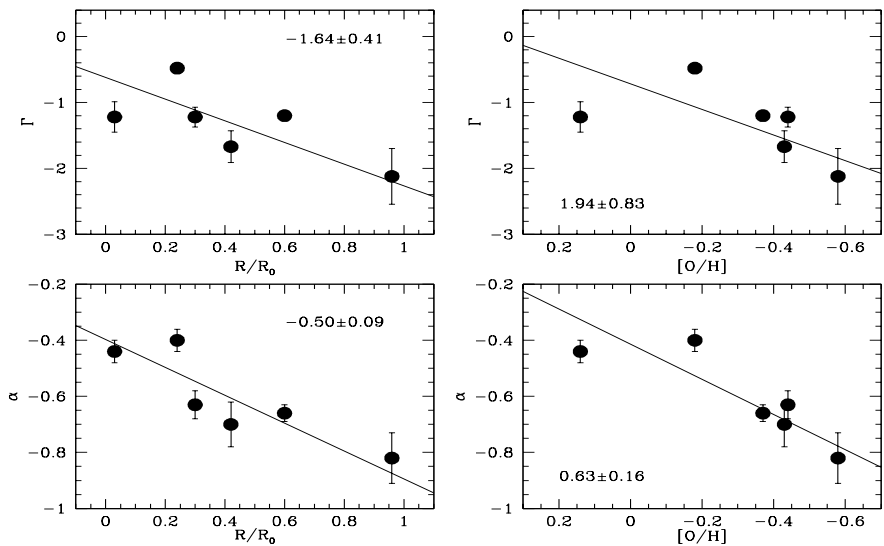,width=13cm,height=7.5cm} }
\caption{
 Relations among $\alpha$ (the slope of the luminosity function), 
$\Gamma$ (the slope of the stellar IMF), 
$[$O/H$]$ (the abundance derived from the HII regions), and
the galactocentric distance of the clusters in M33.
The solid lines represent linear fits to the data. }
\end{figure}


\begin{references}
\reference Dolphin, A. E. 2000, \pasp, 112, 1383
\reference Hunter, D. A., Baum, W. A., O'Neil, E. J.,
   \& Lynds, R. 1996, \apj, 456, 174 
\reference Lejeune, T., \& Schaerer, D. 2001, \aap, 366, 538
\reference Malumuth, E. M., Waller, W. H., \& Parker, J. Wm. 1996, \aj, 111, 1128
\reference Vilchez, J. M., Pagel, B. E.,
   Diaz, A. I., Terlevich, E., \& Edmunds, M. G. 1988, \mnras, 235, 633


\end{references}
\end{document}